%Paper: hep-ph/9503345
%From: anelson@fermi.phys.washington.edu (Ann Nelson)
%Date: Wed, 15 Mar 1995 16:39:02 -0800

\def\np#1#2#3{Nucl. Phys. B{#1} (#2) #3}
\def\pl#1#2#3{Phys. Lett. {#1}B (#2) #3}
\def\prl#1#2#3{Phys. Rev. Lett. {#1} (#2) #3}
\def\physrev#1#2#3{Phys. Rev. {D#1} (#2) #3}

\overfullrule=0pt
%
%     plain-TeX version for world scientific conference proceedings
%
%--------------------------- CUT HERE --------------------------------------
%
%Updated with effect from: 5 Sept 1991
\headline={\ifnum\pageno=1\firstheadline\else
\ifodd\pageno\rightheadline \else\leftheadline\fi\fi}
\def\firstheadline{\hfil}
\def\rightheadline{\hfil}
\def\leftheadline{\hfil}
        \footline={\ifnum\pageno=1\firstfootline\else\otherfootline\fi}
\def\firstfootline{\rm\hss\folio\hss}
\def\otherfootline{\hfil}
\font\tenbf=cmbx10
\font\tenrm=cmr10
\font\tenit=cmti10
\font\elevenbf=cmbx10 scaled\magstep 1
\font\elevenrm=cmr10 scaled\magstep 1
 1

%\TagsOnRight
\nopagenumbers
\line{\hfil }
\vglue 1cm
\hsize=6.0truein
\vsize=8.5truein
\parindent=3pc
\baselineskip=10pt
\centerline{\tenbf CONSEQUENCES OF LOW ENERGY}
\vglue 12pt
\centerline{\tenbf DYNAMICAL SUPERSYMMETRY BREAKING}
%\vglue 0.2cm
%\centerline{\tenbf MANUSCRIPT USING COMPUTER SOFTWARE}
\vglue 5pt
%\centerline{\ninerm (For 20\% Reduction to 8.5 $\times$ 6 in Trim Size)}
\vglue 1.0cm
\centerline{\tenrm ANN E. NELSON }
\baselineskip=13pt
\centerline{\tenit Department of Physics Box 351560}
\baselineskip=12pt
\centerline{\tenit University of Washington, Seattle, WA 98195-1560}
%\vglue 0.3cm
%\centerline{\tenrm and}
%\vglue 0.3cm
%\centerline{\tenrm SECOND AUTHOR'S NAME}
%\centerline{\tenit Group, Company, Address, City, State ZIP/Zone, Country}
\vglue 0.8cm
\centerline{\tenrm ABSTRACT}
\vglue 0.3cm
{\rightskip=3pc
 \leftskip=3pc
 \tenrm\baselineskip=12pt%\parindent=1pc
 \noindent
Relatively simple models can be constructed
in which supersymmetry is dynamically broken at
 energies of $10^5-10^7$ GeV.   Models
of this kind do not suffer from the
naturalness and cosmological
difficulties of conventional supergravity models, and
make   definite predictions for the spectrum of supersymmetric particle masses.
Thus
``Renormalizable Visible Sector Models'' are a  viable alternative to more
conventional approaches. This talk mostly summarizes the results of
reference~1.
\vglue 0.8cm }
\line{\elevenbf 1. Introduction\hfil}
%\medskip
%\line{\elevenit 1.1. Typeset Scripts \hfil}
\medskip
\baselineskip=14pt
\elevenrm
A supersymmetric extension of the Standard Model, with supersymmetry
dynamically
broken by exponentially small nonperturbative effects, provides an attractive
technicolor-like solution to the hierarchy problem, while allowing
the quarks and
leptons to get mass from  Yukawa couplings. Usually we assume that
supersymmetry
is spontaneously broken in a gravitationally coupled ``hidden'' sector

leading to
soft explicit supersymmetry breaking terms in the effective theory. The
resulting
model, known as the Minimal Supersymmetric  Standard Model (MSSM) is
superficially
simple and appealing, but has several theoretically unpleasant features.
\smallskip\noindent\item{1.} The Desert. A fun aspect of supersymmetry
is that it allows us to obtain exact results about strongly interacting gauge
theories. However in the MSSM we have nothing but boring perturbative physics
to
explore below the Planck scale and the interesting dynamics of supersymmetry
breaking
is hidden.
 \noindent\item{2.} The Proliferation of  Parameters. If nonrenormalizable
interactions induced by Planck scale physics have the most general allowed form
 there are  over 100 new parameters required
by soft supersymmetry breaking terms.
 \noindent\item{3.} Naturalness. The Standard Model explains the obseved
absence of baryon and lepton number nonconservation, and the small size of weak
CP
violation and Flavor Changing Neutral Currents (FCNC) in a satisfying way,
without
resorting to speculation about Planck scale physics. In the MSSM the low energy
effective theory provides  no compelling reason for new sources of  violation
 of
these quantum numbers to be absent.
 \noindent\item{4.} The Scale of Supersymmetry Breaking. In the MSSM the
superpartner masses are theoretically not very constrained, and some sort of
conspiracy seems to be required to keep them all out of current experimental
reach.
 \noindent\item{5.} The Mechanism for Dynamical Supersymmetry Breaking (DSB).
In
the MSSM is  the supersymmetry breaking sector is hidden, thus  one doesnt
have to specify it and one can imagine  that it is beautiful. However
explicit hidden sector models  are uncompelling.
 \noindent\item{6.} Cosmology. Most existing hidden sector models of
supersymmetry breaking produce a weak scale gravitino and  weak scale scalars
with
gravitational strength couplings. These typically dominate the energy density
of the
universe until temperatures below 1 keV, which is later than required by
nucleosynthesis$^2$.
\smallskip
An alternative to the MSSM, which ameliorates all of these problems, is to
communicate supersymmetry breaking to the   superpartners via renormalizable
gauge
interactions.
Then the squark, slepton and gaugino masses are calculable from   a
small number of  parameters and the flavor symmetries of gauge interactions
automatically guarantee that the squarks and
sleptons are   sufficiently degenerate
  to prevent FCNC. The trouble-free
cosmology of such models was discussed in ref.~2.
\medskip
%\line{\elevenit 1.2. Section Headings \hfil}
%\smallskip
%\vglue 0.6cm
\line{\elevenbf 2. The Minimal Model of Dynamical Supersymmetry
Breaking\hfil}
\medskip
%\vglue 0.4cm
The simplest known model in which it is possible to obtain DSB in a  limit
where a
reliable calculation can be made has gauge group SU(3)$\times$SU(2)
(``supercolor'')
and matter fields in the
$(3,2)+2(\bar 3,1)+(1,2)$ representation (``supercolored fields'') and the most
general renormalizable
superpotential allowed by the gauge symmetry$^3$. One might think that the
simplest
possible theory would involve this model together with gravitational
transmission of
the information that supersymmetry is broken to ordinary matter, however such
models
have difficulty obtaining sufficient ordinary gaugino masses$^3$. The simplest
way to transmit the
information that supersymmetry is broken is to gauge an additional U(1)
 (``messenger
hypercharge'') which is carried by the supercolored fields and also by some
additional supercolor
singlet ``messenger fields''. Thus we believe that the simplest possible theory
of
supersymmetry breaking involves an   additional  SU(3)$\times$SU(2)$\times$U(1)
gauge group.
Some of the messenger fields   couple to a gauge singlet field ``S'', which
gets an
expectation value and whose F-term becomes nonzero at two loops. S  also
couples to
new quark and lepton superfields with vector-like quantum numbers under the
ordinary
SU(3)$\times$ SU(2)$\times$U(1).
\medskip
\line{\elevenbf 3. Consequences \hfil}
\medskip
Now we can have some fun and calculate the spectrum of ordinary superpartner
masses
induced by ordinary gauge interactions with the new vector-like particles. We
find
that the superpartner masses satisfy to leading order in $F_S$ and gauge
interactions
$$\eqalign{ m_{\rm type\ a\ gaugino}= &k_F^{(a)}{g_a^2
\over 16
\pi^2} {F_S \over S}\cr
  m^2_{\rm squark,\ slepton} = &\sum_a   C_F^{(a)}
\left({g^{(a)2} \over 16 \pi^2}\right)^2{F_S^2 \over S^2}\ ,\cr}$$
where $a$ denotes
the gauge group, $k_F$ is the index of the vector-like representation and $C_F$
 is
the Casimir.  Note that these predictions do not depend on the specific
supersymmetry
breaking sector but are typical of models where gauge interactions communicate
supersymmetry breaking. Other effective supersymmetry breaking terms, such as
trilinear
scalar interactions  as well as other terms which are not soft, can also be
computed and are
small. To this order the Higgs mass squared also comes out positive (and
degenerate
with the left handed sleptons), however at three loops there is the usual
  large
negative contribution induced by the top quark radiative correction$^4$.

Another model independent prediction of low energy supersymmetry breaking is
that the
gravitino is light and is the Lightest Supersymmetric Particle (LSP), with mass
$m_{3/2}=M_s^2/m_P$, where $M_s$ is the supersymmetry breaking scale. Gravitino
cosmology constrains $m_{3/2}$ to be less than 10 keV, to avoid contributing
too much
mass density to the universe$^5$, giving an upper bound on $M_s$ of
$10^7$ GeV. A lower bound on $M_S$ can be obtained by noting that DSB models
generically contain a spontaneously broken U$(1)_R$ symmetry and a Goldstone
boson
known as the R-axion$^6$. The simplest way to give the R-axion an acceptably
large mass$^{6,7}$
is to explicitly break the symmetry via dimension-5 operators suppressed by
$1/m_P$--the mass will be greater than 10 MeV for $M_s > 10^5$ GeV.
The Bino is   typically the next to lightest supersymmetric particle (NLSP),
and
will decay into a gravitino and a photon or $Z$, with lifetime$^8$
$$\tau_{\rm Bino}\sim 8 \pi {M_s^4\over M_{\rm Bino}^5}\ .$$
If supersymmetry is discovered, a search
for the  decay of the NLSP will provide a  lower bound on the supersymmetry
breaking scale, or
evidence for low energy supersymmetry breaking.\medskip
\line{\elevenbf 4. Acknowledgements \hfil}\medskip
This work was supported in part by the DOE under contract \#DE-FG06-91-ER40614
and by a
fellowship  from the Sloan Foundation.
\vglue 0.4cm
\line{\elevenbf 5. References \hfil}\medskip
\noindent\item{1.} M. Dine, A.E. Nelson, Y.Shirman, hep-ph-9408384,
SCIPP-94-21, UW/PT 94-07,
to be published in Phys. Rev. D.
\noindent\item{2.} T. Banks, D.B. Kaplan, A.E. Nelson, \physrev{49}{1994}{779};
B. de Castro, J.A. Casas, F. Quevado, E. Roulet, \pl{318}{1993}{447};
T. Banks, M.
Berkooz, P.J. Steinhardt, hep-th-9501053, RU-94-92.
\noindent\item{3.} I.
Affeck, M. Dine, N. Seiberg, \np{416}{1985}{557}.
\noindent\item{4.} L. Ibanez and G. Ross, \pl{110}{1982}{215}; L.
Alvarez-Gaume, M. Claudson and
M.B. Wise, \np{207}{1982}{96}.
\noindent\item{5.} H. Pagels and J.R. Primack, \prl{48}{1982}{223}.
\noindent\item{6.} A.E. Nelson and N. Seiberg,
\np{416}{1994}{46}.
\noindent\item{7.}  J.
Bagger, E. Poppitz and L. Randall, \np{426}{1994}{3}.
\noindent\item{8.} P. Fayet, \pl{70}{1977}{461}.
\vfill
\eject
\bye